\documentclass[aps,rsi,floatfix,twocolumn,superscriptaddress,footinbib,reprint]{revtex4}

\usepackage{amsmath}
\usepackage{amsfonts}
\usepackage{amssymb}
\usepackage[OT1,T1]{fontenc}
\usepackage[latin1]{inputenc}
\usepackage[english]{babel}
\usepackage[toc,page]{appendix}
\usepackage{verbatim}
\usepackage{multirow}
\usepackage{textcomp}
\usepackage{moreverb}
\usepackage{setspace}

\usepackage{epsf}
\usepackage{graphicx}
\usepackage{graphics}
\usepackage{psfrag}
\usepackage{caption}
\usepackage{subfigure}
\usepackage{placeins}
\usepackage{color}
\usepackage{subfigure}

\usepackage{mathrsfs}
\usepackage{anysize}
\usepackage{amsmath}
\usepackage{amsfonts}
\usepackage{amstext}
\usepackage{amssymb}
\usepackage{bm}
\usepackage[squaren,Gray]{SIunits}

\newcommand{\vor}{Vorono\"\i}

\newcommand{\kar}{von K\'{a}rm\'{a}n}

\newcommand{\ea}{\emph{et al.}}

\begin{document}

\title{Do finite size neutrally buoyant particles cluster?}

\author{L.~Fiabane}
\affiliation{Laboratoire de Physique, ENS de Lyon, UMR CNRS 5672, Université de Lyon, France}
\author{R.~Volk}
\affiliation{Laboratoire de Physique, ENS de Lyon, UMR CNRS 5672, Université de Lyon, France}
\author{J.-F.~Pinton}
\affiliation{Laboratoire de Physique, ENS de Lyon, UMR CNRS 5672, Université de Lyon, France}
\author{R.~Monchaux}
\affiliation{Unité de Mécanique, ENSTA ParisTech, Paris, France}
\author{A.~Cartellier}
\affiliation{LEGI, UMR CNRS 5519, Université Joseph Fourier, Grenoble, France}
\author{M.~Bourgoin}
\affiliation{LEGI, UMR CNRS 5519, Université Joseph Fourier, Grenoble, France}

\date{\today}

\begin{abstract}
We investigate the preferential concentration of particles which are neutrally buoyant but with a diameter significantly larger than the dissipation scale of the carrier flow. 
Such particles are known not to behave as flow tracers (Qureshi \ea, Phys. Re. Lett. 2007) but whether they do cluster or not remains an open question.
For this purpose, we take advantage of a new turbulence generating apparatus, the Lagrangian Exploration Module which produces homogeneous and isotropic turbulence in a closed water flow. 
The flow is seeded with neutrally buoyant particles with diameter $700\,\micro\meter$, corresponding to $4.4$ to $17$ times the turbulent dissipation scale when the rotation frequency of the impellers driving the flow goes from $2\,\hertz$ to $12\,\hertz$, and spanning a range of Stokes numbers from 1.6 to 24.2. 
The spatial structuration of these inclusions is then investigated by a \vor{} tesselation analysis, as recently proposed by Monchaux \ea{} (Phys. Fluids 2010), from images of particle concentration field taken in a laser sheet at the center of the flow. 
Whatever the rotation frequency and subsequently the Reynolds and Stokes numbers, the particles are found not to cluster.
The Stokes number by itself is therefore shown to be an insufficient indicator of the clustering trend in particles laden flows.
\end{abstract}

\maketitle
 
\section{Introduction}
Turbulent flows laden with inertial particles are omnipresent in the industry (chemical reactors, engines, etc.) and the environment, being human-made or not (pollutant dispersion, volcanic or classical cloud formation and dispersion, etc.).
Their study is therefore of great interest and holds many fundamental aspects, issues and limits so far. 
 
One striking feature of these flows is the trend for the particles to concentrate in preferentially sampled regions of the carrier flow.
This has been observed and investigated for long both in experiments~\cite{squires:1991,fessler:1994,bec:2007} and simulations~\cite{yoshimoto:2007,calza:2008}, and it is still widely studied~\cite{bec:2010,monchaux:2010}.
The focus is usually put on small and heavy particles (that is with a high density ratio compared to the fluid), especially in numerical studies.
Because of their high specific density, the dynamics of such small and heavy inertial particles deviates from that of the carrier flow. 
Clustering phenomena are then one of the many manifestations of this departure from tracer behavior.
Some other studies were conducted for light particles as well, exhibiting the same trend to cluster but with different cluster geometries~\cite{calza:2008}.
Finally tracers (ought to be both neutrally buoyant and much smaller than the dissipative scale of the carrier flow) are usually used to characterize the flow dynamics.
The case of finite size neutrally buoyant particles, however, has never been treated to our knowledge in the context of preferential concentration phenomenon. 
Such finite size particles (with a diameter significantly larger than the dissipation scale of the carrier flows) are known experimentally~\cite{qureshi:2007} and numerically~\cite{bec:2010}, to differ from tracers. 
However, existing studies have focused on the dynamics of isolated particles, but not on the spatial structuration of laden flows. 
Whether they cluster or not remains an open question.
In order to study the preferential concentration of finite size - neutrally buoyant particles, we perform a \vor{} tesselation analysis.
This technique, recently introduced for the investigation of preferential concentration in two~\cite{monchaux:2010} or three~\cite{tagawa:2012} dimensions, is particularly efficient and robust to diagnose and quantify clustering phenomenon.

The article is organized as follows. In section \ref{sec:expe} we describe the experimental setup and the data processing used to carry this investigation. Section \ref{sec:results} describes the results on preferential concentration of finite size - neutrally buoyant particles. We finish with a brief discussion and conclusions (section \ref{sec:discussion}).

\section{Experimental setup and postprocessing}
\label{sec:expe}

\subsection{Turbulent flow generation}

In order to study the behavior of neutrally buoyant particles, a straightforward solution is to create a turbulent water flow, with the particle density matching that of water (neutrally buoyant particles can be obtained in air, for instance with soap bubbles inflated with Helium~\cite{qureshi:2007, qureshi:2008} ; however while it is easy to produce such bubbles individually, dense seeding with numerous particles is difficult).
Many experimental apparatuses creating turbulent water flow with small mean flow velocities exist, the best known example being the \kar{} flow.
This \kar{} flow is a high Reynolds number turbulent flow created between two counter-rotating disks.
Near the center of the apparatus, the mean velocities are much weaker than the fluctuations. 
This is of particular interest for the investigation of particles in turbulence as the mean central stagnation point eases the acquisition of long particle trajectories, what has made the \kar{} flow the natural choice for several pioneering experiments on Lagrangian particle tracking~\cite{laporta:2001,mordant:2001}. However, \kar{} flows exhibit statistical inhomogeneity and anisotropy that may render the interpretation of the results difficult, in particular when it comes to discriminate between effects associated to the large anisotropic structures and the small turbulent scales.

\begin{figure}[t]
\centering
\subfigure[]{\includegraphics[width=0.25\textwidth]{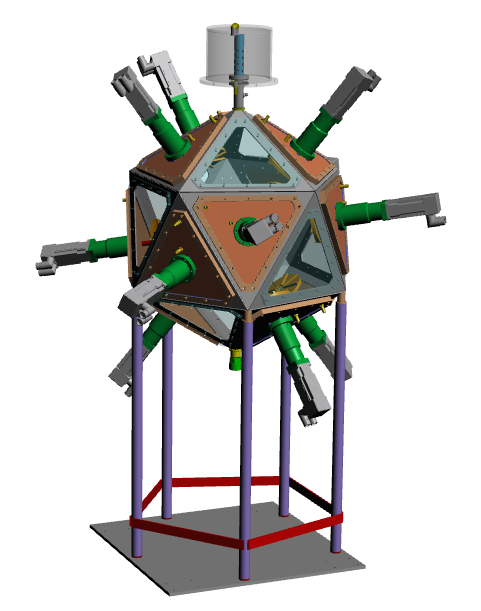}}
\subfigure[]{\includegraphics[width=0.4\textwidth]{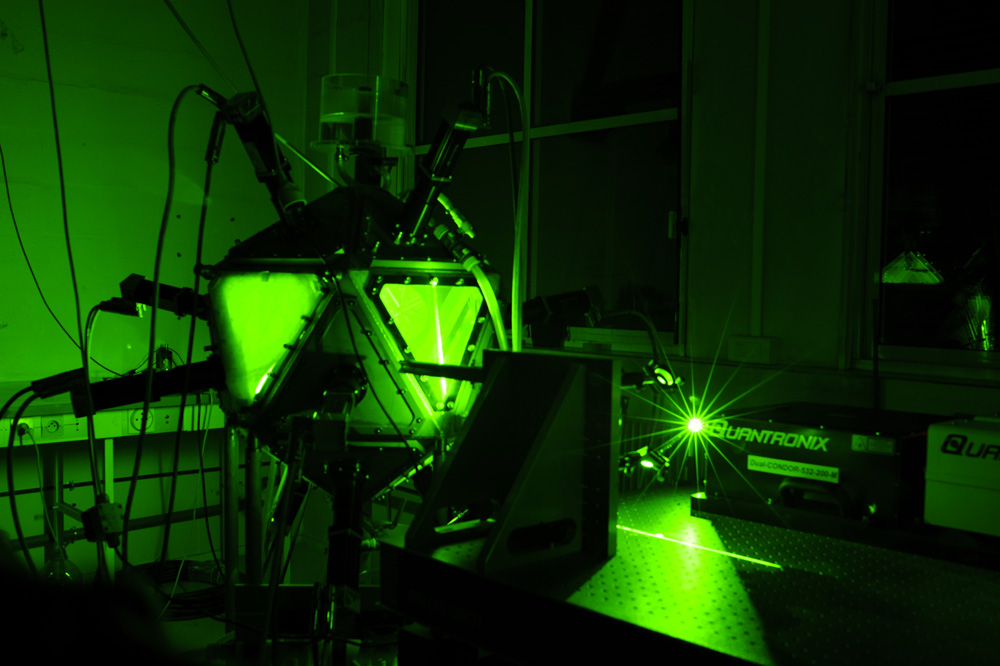}}
\caption{(a)~CAD rendering of the LEM \; (b)~LEM and Nd:YAG laser \emph{in situ}}
\label{fig:lem}
\end{figure}

Experiments in this study are conducted in a new turbulence generating apparatus, the Lagrangian Exploration Module (LEM), developed in collaboration between the Laboratoire de Physique of the \'Ecole Nationale Sup\'erieure de Lyon and the Max Planck Institute of Dynamics and Self-Organization in G\"ottingen.
The LEM produces turbulence in a closed water flow driven by twelve impellers evenly distributed on twelve of the twenty faces of an icosahedral vessel (see Figure~\ref{fig:lem}).
The length of the edges of the icosahedron is $40\,\centi\meter$, giving a volume of $140\,\liter$ water.
The twelve impellers can be independently driven. In the present work, all impellers are used simultaneously and rotate at the same constant frequency $f$, which can be increased up to $12.5\,\hertz$, with the constraint that impellers in front of each other counter-rotate. 
This has been shown to achieve a statistically homogeneous and isotropic flow with almost zero mean velocity in a central region of the device of order $10\times10\times10\,\centi\meter^3$~\cite{zimmermann:2010}.

For the present setup, turbulence has been characterized with two-dimensional particle image velocimetry (PIV) using LaVision device on a $15\,\centi\meter \times 10\,\centi\meter$ plane in the center of the LEM. 
It confirms a nearly homogeneous and isotropic turbulence in a sphere of nearly $8\,\centi\meter$ diameter at the center of the LEM. 
The evolution of the turbulence characteristics (defined below) for the explored range of rotation frequencies is given in Table~\ref{tab:characteristics}.
\begin{table*}[tb]
\begin{ruledtabular}
\begin{tabular}{cccccccc}
$f$ & $u'$ & $\varepsilon$ & \multirow{2}{*}{$R_{\lambda}$} & $L$ & $\eta$ & $\tau_{\eta}$ & \multirow{2}{*}{$St$}\\
($\hertz$) & ($\centi\meter\per\second$) & ($\squaren\meter\per\second^3$) &  & ($\centi\meter$) & ($\micro\meter$) & ($\milli\second$) & \\
\hline 
2 & 4 & 0.0016 & 160 & 4.0 & $158$ & $24.9$ & 1.6\\
4 & 8 & 0.0144 & 210 & 3.6 & 91 & 8.3 & 4.7\\
6 & 12 & 0.0611 & 260 & 2.9 & 64 & 4.0 & 8.6\\
8 & 17 & 0.1086 & 310 & 4.5 & 55 & 3.0 & 13.2\\
10 & 22 & 0.2087 & 360 & 5.1 & 47 & 2.2 & 18.4\\
12 & 26 & 0.3518 & 395 & 5.0 & 41 & 1.7 & 24.2
\end{tabular}
\end{ruledtabular}
\caption{Evolution of the turbulence characteristics in the LEM with the rotation frequency $f$ of the twelve impellers.
$u'$:~the fluctuation velocity, averaged over the entire measurement plane.
$\varepsilon$:~the energy dissipation rate, measured from the inertial range scaling of the Eulerian velocity structure functions. 
$R_\lambda$:~the Taylor micro-scale Reynolds number, calculated from Eq.~(\ref{eq:Rlambda}).
$L$:~the integral length scale.
$\eta \equiv (\nu^3/\varepsilon)^{1/4}$ and $\tau_\eta \equiv (\nu/\varepsilon)^{1/2}$:~the Kolmogorov length and time scales of the flow, respectively.
$St$:~the Stokes number of particles as defined in Eq.~(\ref{eq:stokes}).}
\label{tab:characteristics}
\end{table*}
The fluctuating velocities are averaged over the entire illuminated plane as:
$u' \equiv \tfrac{\langle u_{x,rms} \rangle + \langle u_{y,rms} \rangle}{2}$.
We estimate the energy dissipation rate $\varepsilon$ from scalings of the velocity structure functions in the inertial range (see, e.g.~\cite{pope:2000}):
\begin{align}
\varepsilon = & \frac{1}{r} \bigg[ \frac{D_{LL}(r)}{C_2} \bigg]^{3/2} , \\
\label{eq:compensatedDLL}
\varepsilon = & \frac{1}{r} \bigg[ \frac{3 D_{NN}(r)}{4 C_2} \bigg]^{3/2},  
\end{align}
where $D_{LL}(r)$ and $D_{NN}(r)$ are the second-order longitudinal and transverse structure functions.
The constant $C_2$ is set to $C_2 = 2.1$ as obtained from a compilation of data in various turbulent flows~\cite{sreenivasan:1995}. 
Moreover, we have checked that the isotropic conditions $D_{NN}=4/3\,D_{LL}$ holds over the entire range of resolved scales what confirms the good isotropy properties of the LEM. The estimations of $\varepsilon$ from the transverse and the longitudinal scaling are in excellent agreement.
The integral length scale $L$ is defined by:
\begin{equation}
L \equiv \int_0^\infty D_{LL}(r) \, dr.
\label{eq:Lint}
\end{equation}
Finally the Reynolds numbers based on the Taylor microscale (simply referred to as the Reynolds number later on) is defined as:
\begin{equation}
R_\lambda \equiv \sqrt{15\frac{u' \,L }{\nu}} = \sqrt{15 \frac{\varepsilon^{1/3}L^{4/3}}{\nu}} = \sqrt{\frac{15 u'^4}{\nu \varepsilon}} .
\label{eq:Rlambda}
\end{equation}
When the rotation velocity is changed from 2 to 12 $\hertz$, the associated Reynolds numbers based on the Taylor microscale varies from 160 to 395 in the center of the LEM.

\subsection{Particles characteristics}

Regarding the particles used, one of the main goals of our study is to explore the behavior of finite size neutrally buoyant particles.
More precisely the particles used must be neutrally buoyant in water, meaning their density $\rho_\text{particle}$ must be close to $\rho_\text{water}$.
Their size is also of importance: not so small that they would behave as tracers and follow the carrier flow dynamics, but also small enough so that the flow dynamics is not much altered. 

In concrete terms, we chose polystyrene particles of diameter $d=700\,\micro\meter$ -- corresponding to $4.5$ to $17$ times the Kolmogorov length scale $\eta$, depending on the turbulence energy dissipation rate $\varepsilon$.
This large range of ratio between the particles diameter and the Kolmogorov length scale allows us to study what can be seen as tracers ($d/\eta\approx 4.5$) on the one hand, and inertial particles ($d/\eta > 5$) on the other hand~\cite{brown:2009,volk:2011}.
The particles density $\rho_\text{particle}$ has been adjusted so that the ratio $\Gamma=\tfrac{\rho_\text{particle}}{\rho_\text{water}}$ is  $1\leq \Gamma \leq 1.015$. 
These particles are obtained from small expandable polystyrene particles with original density of order $1.05$. 
These particles are irreversibly expanded by a moderate heating so that as they expand their density decreases. 
In the expansion process particles whose density matches as close as possible that of water are then selected and sieved.

Particles interacting with a turbulent flow are commonly characterized by their Stokes number, a dimensionless number which quantifies the ratio between the particle viscous relaxation time and a typical time scale of the flow. The latter is generally chosen to be the Kolmogorov time scale:

\begin{equation} \label{eq:stokes}
St \equiv \frac{\tau_p}{\tau_{\eta}}=\left( \frac{d}{\eta} \right)^2 \frac{1+2\Gamma}{36},
\end{equation}
where $\tau_p$ is the particle viscous relaxation time, $\tau_{\eta}$  the dissipation time scale of the carrier flow and $\eta$ its dissipation length scale.
The Stokes number is usually used as the key --and often the only-- parameter to characterize particle dynamics in turbulence.
This is highly motivated by the simplicity of Stokesian models for modeling and numerical simulation purposes, where the dominant force acting on the particle is simply taken as the drag due to the difference between the particle  velocity $v$ and the fluid velocity $u$:
\begin{equation}\label{eq:stokesian}
\frac{d\vec{v}}{dt}=\frac{1}{\tau_p}(\vec{u}-\vec{v}),
\end{equation}
the only explicit relevant parameter for the particles being then the viscous response time.
This minimal Stokesian model, whose validity can only be warrantied as an approximation of Maxey \& Riley and Gatignol equation~\cite{maxey:1983,gatignol:1983} for the case of small inertial particles much heavier than the fluid, is commonly used in numerical simulations (both DNS and kinematic simulations) investigating the turbulent dynamics of inertial particles. 
An important result to be emphasized in the context of the present study is the observed dependence of preferential concentration with the Stokes number. 
More specifically, Stokesian models suggest that particles with non-vanishing Stokes number tend to exhibit preferential concentration.
Further clustering and segregation is maximal for particles whose Stokes number is of order unity~\cite{bec:2007,coleman:2009}. 
This trend is supported by experimental measurements of the concentration field of small inertial particles~\cite{monchaux:2010}.

For the particles investigated in the present work, the Stokes number as defined in Eq.~(\ref{eq:stokes}) spans a wide range from 1.6 to 24 as shown in Table~\ref{tab:characteristics}. 
Though our particles are finite size, and hence Stokesian approximation (Eq.~\ref{eq:stokesian}) is not expected to hold by itself, the question of the influence of the Stokes number on the spatial distribution of such finite size particles is still of relevant interest. 
In the present study, the particle concentration field has been investigated as a function of their Stokes number. 
It has to be noted that as the particle diameter and density are kept constant, the Stokes number is varied by tuning the flow dissipation time scale in Eq.~(\ref{eq:stokes}). 
Therefore, it cannot be varied independently of the Reynolds number of the carrier flow.

\subsection{Acquisitions and postprocessing}

\subsubsection{Particles detection}
Acquisitions are performed using 12 bits digital imaging at a resolution of 2400$\times$1800 pixels corresponding to a $15\,\centi\meter \times 10\,\centi\meter$ visualization window in the center of the LEM. 
Images are recorded with a Phantom V10 camera (Vision Research Inc.) operated at a low repetition rate of $2.5\,\hertz$ (note that we only address here the question of particle spatial distribution and we do not aim at tracking particle dynamics, what would have required a much higher repetition rate).
The visualization window is illuminated by a $100\,\watt$ pulsed Nd:YAG laser (Condor Serial, Quantronix) synchronized with the camera, creating a light sheet with millimetric thickness.
The camera is mounted with a $90\,\milli\meter$ macro lens (Tamron) through a Scheimpflug mount to compensate for the depth of field effects resulting from the angle between the camera and the laser sheet.
Each experiment consists in 2000 uncorrelated images acquired in nearly $15\,\minute$ for a fixed rotation frequency for each motor and a constant concentration of polystyrene in the LEM.
We identify the particles on the images as local maxima with light intensity higher than a threshold, assuming in a first approximation that all the particles illuminated in the laser sheet belong to one plane.
The center of the particles is determined as the center of mass of all the pixels surrounding one local maxima.
We have checked that changing slightly the threshold value does not significantly impact the number of detected particles ;
essentially because the contrast between the light diffused by the particles and the background is very strong due to the large size of the particles.
At the working seeding density, the average number of detected particles is of order 100.
No diminution of the number of detected particles is observed from the beginning to the end of an experiment.
This indicates a good stationarity of seeding concentration as expected for non-settling neutrally buoyant particles.

\begin{figure}[tb]
\centering
\subfigure[]{\includegraphics[width=0.48\textwidth]{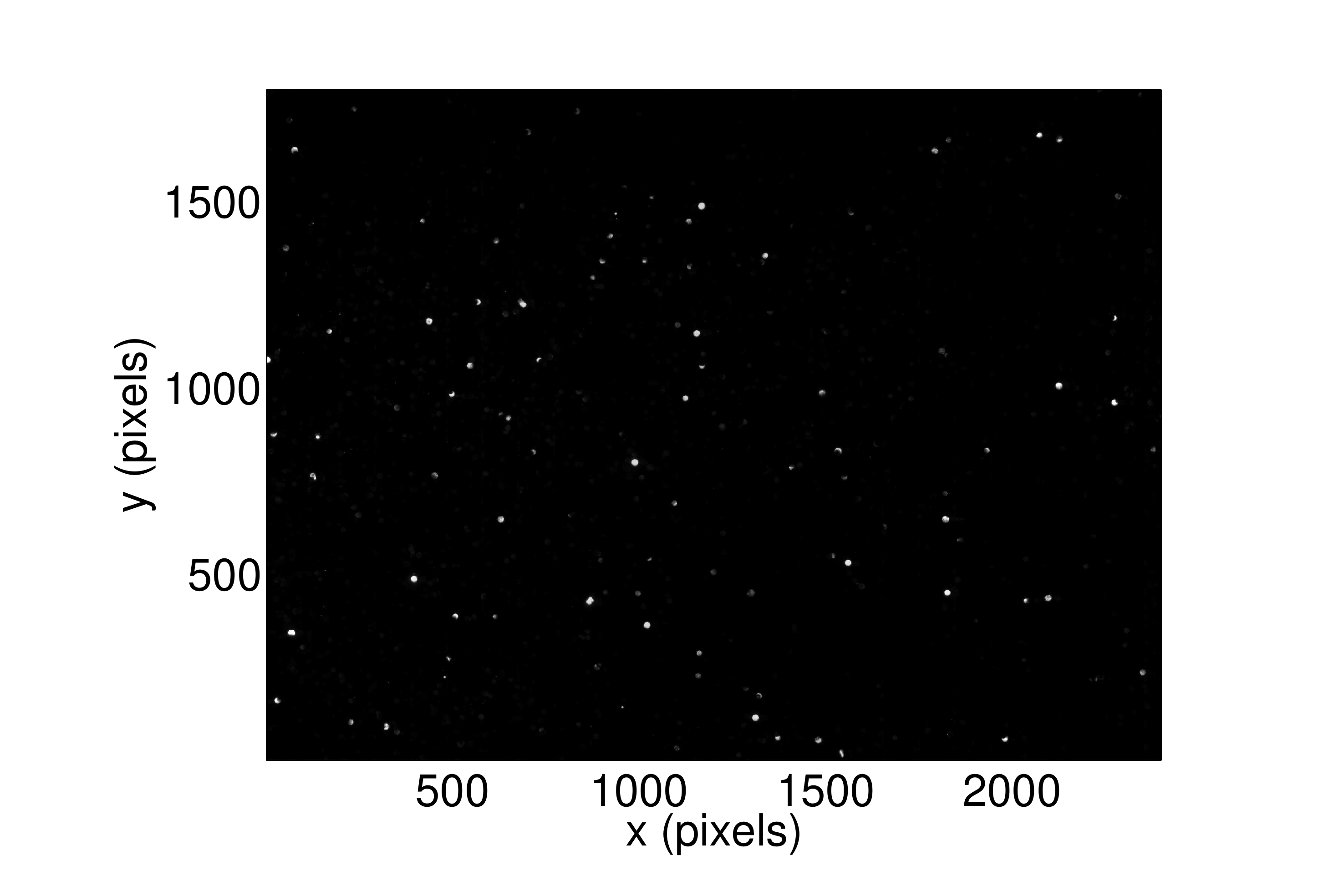}}
\subfigure[]{\includegraphics[width=0.48\textwidth]{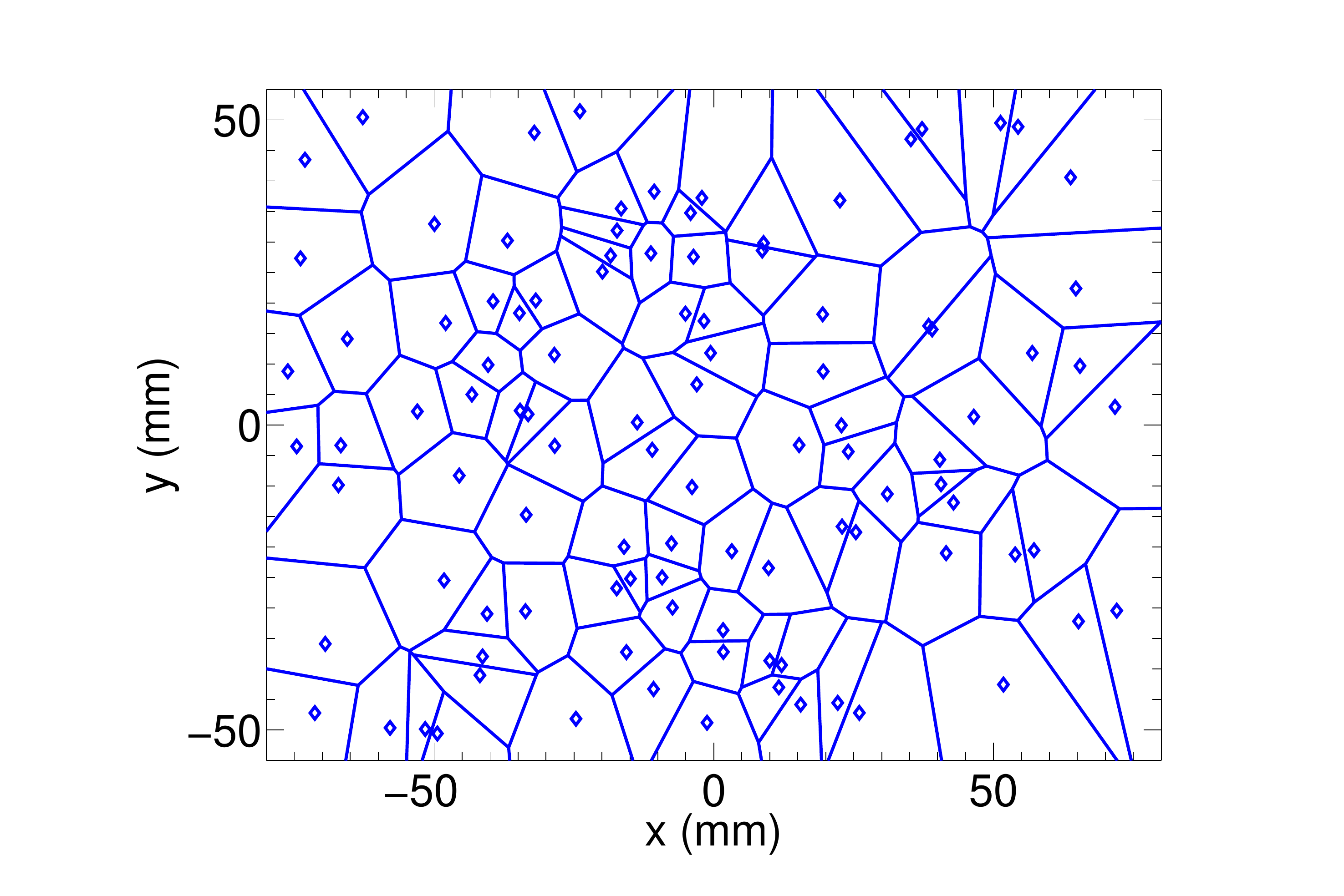}}
\caption{(a) Typical raw acquired image. \; (b)~Particles located in this image and the associated \vor{} diagram.}
\label{fig:tesselation}
\end{figure}

\subsubsection{\vor{} tesselation study}
In order to study the particle concentration field, we use \vor{} diagrams.
A raw acquired image, and the detected particles and the associated \vor{} diagram, are provided in Figures~\ref{fig:tesselation}(a) and \ref{fig:tesselation}(b) respectively.
Such a \vor{} analysis has been recently introduced for the investigation of preferential concentration of small water droplets in a turbulent flow of air~\cite{monchaux:2010} , and was shown to be particularly efficient and robust to diagnose and quantify clustering phenomenon.
These \vor{} diagrams give a tesselation of a two-dimensional space where each cell of the tesselation is linked to one seed (one detected particle in the present case) in such a way that all the points of the cell are closer to the associated particle than to any other particle.
It appears that the area of a \vor{} cell is the inverse of the local concentration of particles.
Studying \vor{} area field is thus equivalent to studying local concentration fields.
The choice of the \vor{} method was made for two main reasons.
First because unlike other methods, using for instance box counting as a tool to study particles concentration~\cite{fessler:1994}, this technique gives a measure of the local concentration field at interparticle length scale, meaning the measure does not depend on the field size nor on an extrinsic length scale choice.
Second, this method is numerically very efficient regarding the number of particles we have to process (several hundreds per image).
To compare the results of different experiments made with different amount of detected particles per image, one has to find a normalization.
This is achieved using the average \vor{} area $\bar{A}$ defined as the mean particles concentration inverse, which does not depend on the spatial organization of the particles.
Therefore, we focus on the distribution of the normalized \vor{} area $\mathcal{V}\equiv A/\bar{A}$ in the rest of the study.

\section{Results on the preferential concentration} 
\label{sec:results}

Clustering properties can be investigated and quantified by comparing the probability distribution function (PDF) of \vor{} cell areas in the experiment to that of \vor{} cell areas of a synthetic random Poisson process. 
It has to be noted that surprisingly, no analytical form is known for the PDF of \vor{} cell areas of such a Poisson process, though its form is known to be well approximated by a Gamma distribution~\cite{ferenc:2007}.
As a random Poisson process reference, we use here the compact analytical expression involving the space dimension as a single parameter proposed by~\cite{ferenc:2007}.
In two dimensions it reads:
\begin{equation}\label{eq:gammafit}
f(y)=C y^{5/2} \exp(-7y/2),
\end{equation}
with the constant C=24.1358 in our case.

\begin{figure}[t]
\centering
\includegraphics[width=0.48\textwidth]{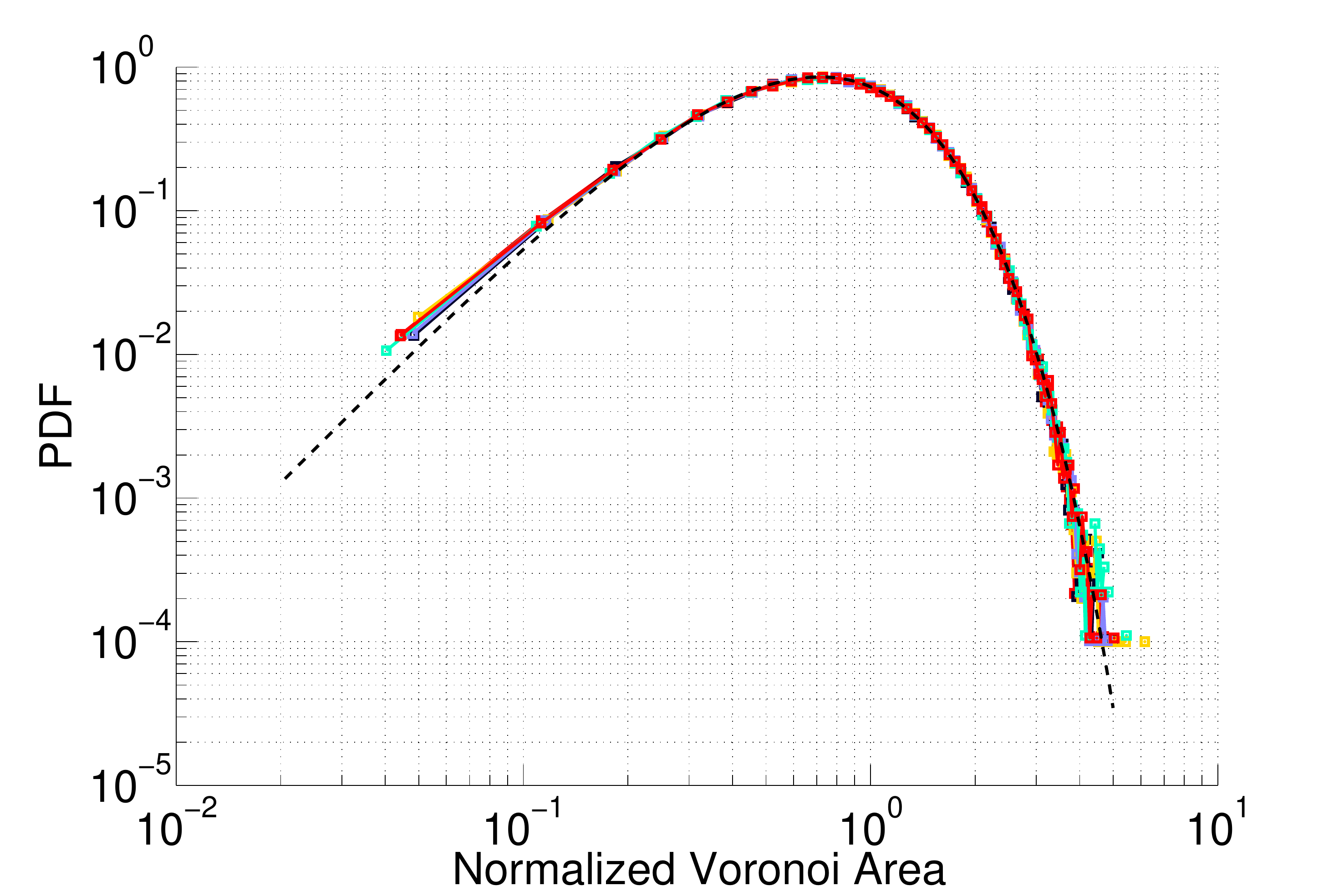}
\caption{Superposition of the normalized \vor{} area ($\mathcal{V}\equiv A/\bar{A}$) PDFs for six experiments with varying Stokes number (colored lines, each PDF is calculated from 2000 instantaneous fields) and a Gamma distribution (black dashed line).}
\label{fig:voro_pdf}
\end{figure}

The PDFs of \vor{} cell areas for the different experiments described in Table~\ref{tab:characteristics} are plotted in Figure~\ref{fig:voro_pdf}. 
As it can be seen, all the PDFs collapse within statistical convergence error bars. 
No systematic trend is visible with the different experimental configurations. 
We have superimposed on the same plot a Gamma distribution. 
Interestingly we find that the PDFs and the Gamma distribution also collapse, meaning the particles do not exhibit any preferential concentration (or clustering) whatever the Stokes number.

\begin{table*}[t]
\begin{ruledtabular}
\begin{tabular}{ccccccc}
$St$ & 1.6 & 4.7 & 8.6 & 13.2 & 18.4 & 24.2 \\
$\sigma_\mathcal{V}$ & 0.532 & 0.530 & 0.529 & 0.530 & 0.535 & 0.533\\
\multicolumn{7}{c}{$\sigma_{\mathcal{V}}^\text{RPP} \approx 0.53$}
\end{tabular}
\end{ruledtabular}
\caption{Standard deviations of the normalized \vor{} areas $\sigma_\mathcal{V}=\sqrt{\langle \mathcal{V}^2 \rangle-1}$ for different Stokes numbers, to be compared with the standard deviation of the normalized \vor{} area corresponding to a random Poisson process $\sigma_\mathcal{V}^\text{RPP}$.}
\label{tab:sigma}
\end{table*}

This result can be further quantified using the standard deviation of the \vor{} areas.
The standard deviations of the normalized \vor{} areas are given in Table~\ref{tab:sigma}.
The experimental standard deviations are all very close to 0.53, the standard deviation of the normalized \vor{} area corresponding to a random Poisson process.
This reveals again the tendency of particles to distribute in a random way.

\section{Conclusion and discussion on the Stokes number}
\label{sec:discussion}

We have investigated in this study the preferential concentration of finite size, neutrally buoyant particles using \vor{} diagrams. 
By varying the Reynolds number in the closed water flow, we have been able to vary the Stokes number (the density and the diameter of the particles being kept constant).
Although such particles are known to have a different dynamics from that of the flow, we have found no preferential concentration in the spatial structuration of these inclusions regardless the Stokes number.

This result is contrary with most of the studies describing preferential concentration of particles as a result of inertial effects.
As mentioned previously, these studies usually account the deviation of the particles dynamics for inertial effects due to the difference in density between the particles and the fluid (hence the name \emph{inertial particles}).
The clustering is then regarded as an effect of the inertia of the particles, and the model used to quantify this preferential concentration usually involves the Stokes number only.

The present study, however, while dealing with particles whose Stokes number is high (above unity) and varies, does not exhibit any clustering.
It definitely shows that the Stokes number by itself is not a sufficient indicator of the clustering trend in particles laden flows.
In our case, we suspect that the pressure distribution at the surface of the particles must be preponderant in the dynamics of the particles.

In future studies we plan to investigate the influence of the flow anisotropy on the spatial structuration of the particles, thanks to the versatility of the LEM and its twelve independently driven impellers.
We will also carry out the same kind of experiments with heavy particles in order to study the impact of the particles sizes on the clustering phenomena and on their geometries, and the influence of anisotropy.

\begin{acknowledgments}
We thank R. Zimmermann for stimulating discussions about the LEM.
This work was partially supported by the ANR Blanc DSPET and the French Government FUI Program PATVAX.
\end{acknowledgments}

\bibliography{biblio}

\end{document}